\documentclass[aps,prb,twocolumn,letterpaper,superscriptaddress,showpacs]{revtex4-1}
\usepackage{keyval}%
\usepackage{graphicx}
\usepackage{dcolumn}
\usepackage{bm}
\usepackage{color}
\usepackage{amsmath,amssymb}
\usepackage{soul}
\usepackage{setspace}

\begin{document}

\title{Realization of intrinsically broken Dirac cones in graphene via the momentum-resolved electronic band structure}
\author{Chi-Cheng Lee}
\affiliation{Institute for Solid State Physics, The University of Tokyo, 5-1-5 Kashiwanoha, Kashiwa, Chiba 277-8581, Japan}%
\author{Masahiro Fukuda}
\affiliation{Institute for Solid State Physics, The University of Tokyo, 5-1-5 Kashiwanoha, Kashiwa, Chiba 277-8581, Japan}%
\author{Yung-Ting Lee}
\affiliation{Institute for Solid State Physics, The University of Tokyo, 5-1-5 Kashiwanoha, Kashiwa, Chiba 277-8581, Japan}%
\author{Taisuke Ozaki}
\affiliation{Institute for Solid State Physics, The University of Tokyo, 5-1-5 Kashiwanoha, Kashiwa, Chiba 277-8581, Japan}%
\date{\today}

\begin{abstract}
A way to represent the band structure that distinguishes between energy-momentum and energy-crystal momentum relationships is proposed upon the
band-unfolding concept. This momentum-resolved band structure offers better understanding of the physical processes requiring the information of wave functions 
in momentum space and provides a good description of angle-resolved photoelectron spectroscopy (ARPES) spectra together with a still informative band structure.  
Following this approach, we demonstrate that Dirac cones in graphene are intrinsically broken in momentum space and can be described by a conceptual unit cell
smaller than the primitive unit cell. This hidden degree of freedom
can be measured by ARPES experiments as missing weight that is retrievable by probing the chirality and Berry phases by linearly and 
circularly polarized light. Having the energy-momentum relationship, we provide alternative understanding of the retrieved momentum intensity, 
that is, the retrieved momentum intensity is assisted with the properties of final states, not from the Dirac cones directly.  
The revealed broken Dirac cones and momenta supplied by the lattice give interesting ingredients 
for designing advanced nanodevices. 
\end{abstract}
  
\maketitle

\section{Introduction}
\label{sec:introduction}

A noticeable amount of research has followed the passionate pursuit of realization of Dirac-fermion-related science and technology in real materials,
such as the topological insulators\cite{RevModPhys.82.3045}. Graphene is no doubt one of the representatives of hosting massless Dirac fermions 
in the class with negligible spin-orbit coupling\cite{RevModPhys.81.109}.
Among many findings in graphene, one intriguing issue is the incomplete Dirac cone observed in the angle-resolved photoelectron spectroscopy (ARPES) spectra,
where a horseshoe pattern in the constant-energy contour has been identified\cite{PhysRevB.51.13614,bostwick2007quasiparticle,PhysRevB.77.195403,
PhysRevB.83.121408,PhysRevB.84.125422,PhysRevLett.107.166803,
gierz2012graphene,hwang2014angle,hwang2015tight,PhysRevB.92.195148,puschnig2015simulation,moser2016experimentalist}. 
Since the energy dispersion of Dirac cones in graphene has been frequently confirmed by tight-binding-model 
and first-principles calculations in the periodic-zone scheme\cite{RevModPhys.81.109}, 
researchers have great enthusiasm for retrieving the missing components of Dirac cones in the momentum space explored by ARPES experiments
\cite{PhysRevB.51.13614,bostwick2007quasiparticle,PhysRevB.77.195403,
PhysRevB.83.121408,PhysRevB.84.125422,PhysRevLett.107.166803,gierz2012graphene,hwang2014angle,hwang2015tight,PhysRevB.92.195148,puschnig2015simulation,moser2016experimentalist}. 
By focusing on the effect of polarization, all of the different pieces of the constant-energy contour become measurable using linearly or circularly polarized light, resulting
from the interplay with the chirality and Berry phases\cite{PhysRevB.83.121408,PhysRevB.84.125422,PhysRevLett.107.166803}. 
While such a realization of complete Dirac cones that reveals the chirality of massless Dirac fermions in graphene described by the periodic-zone scheme is profound, 
an alternative picture beyond the periodic-zone scheme could provide more complementary ingredients to current understanding. 
Moreover, the revealed circular dichroism reflecting the chirality of massless Dirac fermions in graphene cannot be easily reproduced using similar circularly polarized light 
and the applied photon energy seems to play an important role for the observed spectra\cite{gierz2012graphene,hwang2014angle,hwang2015tight}. 

Current emphases of the resulting missing spectral weight are quite diverse, for example, on the property of initial states, the interference between photoexcited electrons, 
the effect of photon polarization, and the choice of the final states\cite{PhysRevB.51.13614,bostwick2007quasiparticle,PhysRevB.77.195403,
PhysRevB.83.121408,PhysRevB.84.125422,PhysRevLett.107.166803,gierz2012graphene,hwang2014angle,hwang2015tight,PhysRevB.92.195148,puschnig2015simulation,moser2016experimentalist}. 
Indeed, ARPES measurement provides a way to access the momentum and binding energy 
of an electron in the Bloch state $|\vec{k}n\rangle$, but a realistic description of ARPES spectra is complicated requiring knowledge of many-body eigenstates, 
multiple scattering, surface effects, and so forth\cite{williams1977direct,RevModPhys.75.473,puschnig2015simulation,moser2016experimentalist}. On the other hand, recently proposed 
band-unfolding methods suggest a different way to understand ARPES spectra from theoretical calculations\cite{PhysRevB.71.115215,PhysRevLett.104.216401,
PhysRevLett.104.236403,PhysRevB.85.085201,PhysRevB.87.085322,PhysRevB.87.245430,lee2013unfolding,huang2014general,PhysRevB.90.115202,PhysRevB.90.195121}. 
It has been found that the ARPES spectra can be more easily compared by representing the bands
in a conceptual Brillouin zone beyond what the geometry can offer\cite{PhysRevLett.107.257001,PhysRevB.90.075422}. 
In fact, the representation of one-carbon-atom zone has been shown to be able to reproduce the incomplete Dirac cones\cite{lin2013create}.
This type of Dirac cones also exists in silicene, germanene, stanene, and other similar materials\cite{balendhran2015elemental}. 
It is then interesting to demonstrate why such a choice can give good agreement with experiments and to address the intrinsic nature of the extensively explored Dirac cones in graphene.

In this study, we provide a comprehensive picture of the broken Dirac cones in graphene and present a substantial clarification for 
understanding the ARPES spectra by the momentum-resolved electronic band structure upon the concept of band unfolding.
We take the opportunity to revisit the band-unfolding methods using plane waves\cite{PhysRevLett.104.236403,PhysRevB.85.085201} and address 
the advantage of performing such unfolding using atomic-like orbitals{\cite{Ozaki}. 
The unfolded spectral weight and the ARPES spectra are then bridged via the momentum distribution of initial states
based on simple basics of quantum mechanics. A general guidance of how to choose a good \textit{unit cell} for performing unfolding is also given.
By showing the Fourier components of massless Dirac fermions in graphene, it is clear that the one-carbon-atom unit cell gives the best description 
of the ARPES spectra together with a still informative band structure. The revealed broken Dirac cones in graphene,  
without considering any photon-excitation processes, nail down a simple but convincing alternative mechanism for explaining the retrieved components of Dirac cones 
in the ARPES measurements and suggest that those retrieved momenta, which are not shown in the unfolded weight, do not belong to the massless Dirac fermions. 

The paper is organized as follows. The band-unfolding method using plane waves and a simplified description of ARPES spectra are discussed in Sec.~\ref{sec:unfolding}
and Sec.~\ref{sec:arpes}, respectively. The connection between the unfolded weight from first-principles calculations and the measurable ARPES intensity is given in Sec.~\ref{sec:connection}. 
In Sec.~\ref{sec:graphene} the broken Dirac cones in graphene revealed in the momentum-resolved electronic band structure are discussed. The conclusion is drawn in Sec.~\ref{sec:conclusion}.

\section{Unfolding band structure}
\label{sec:unfolding}

The picture of independent electrons experiencing periodic potentials described by Bloch theorem is commonly adopted as the first step in
understanding condensed matter\cite{Ashcroft}. The rule of thumb in this approach is the conservation of crystal momentum 
ensuring that an energy eigenstate $|\vec{k}n\rangle$, denoted by the crystal momentum $\vec{k}$ and the band index $n$, 
is equivalent to $|\vec{k}+\vec{G},n\rangle$, where $\vec{G}$ is a reciprocal lattice vector.
As a result, the band structure can be discussed in the first Brillouin zone.
However, the \textit{momentum} distribution of $|\vec{k}n\rangle$ could vary at different plane wave basis, denoted as $|\vec{k}+\vec{G}\rangle$. 
Since $|\vec{k}n\rangle$ can only be expanded by plane waves associated with $\vec{k}+\vec{G}$ vectors, 
one can mathematically redefine the $\vec{k}+\vec{G}$ vectors with a different set of $\vec{G}$ vectors whose reciprocal 
primitive vectors may be longer than the original ones. To complete the summation with the new periodicity, which will be labeled by $\vec{G}^\prime$, 
new $\vec{k}^\prime$ needs to be introduced for $\vec{k}+\vec{G}$ shorter than the new periodicity: 
\begin{align}
|\vec{k}n\rangle & = \sum_{\vec{G}} C^n_{\vec{k}+\vec{G}}|\vec{k}+\vec{G}\rangle \nonumber  \\
                 & = \sum_{\vec{k}^\prime}\sum_{\vec{G}^\prime} C^n_{\vec{k}^\prime+\vec{G}^\prime}|\vec{k}^\prime+\vec{G^\prime}\rangle. 
\label{eq:regroup}
\end{align}
Physically, the $\vec{k}^\prime$ can be considered as the crystal momentum belonging to the first Brillouin zone of a conceptual unit cell
whose reciprocal lattice vectors are the $\vec{G}^\prime$ vectors. Note that $\vec{k}^\prime$ equals $\vec{k}$ plus a $\vec{G}$ vector. 
The conceptual cell should be properly chosen to be commensurate with the studied system. 
It is worth mentioning that the conceptual unit cell can be smaller than the primitive unit cell defined by the geometrical structure. 
How to choose the conceptual unit cell is the topic of unfolding band structure. 

To find the represented intensity of $|\vec{k}n\rangle$ in the conceptual Brillouin zone, we introduce a new ket vector $\parallel\vec{k}^\prime n\gg$
using thicker symbols, such as ``$\parallel$'' and ``$\gg$'': 
\begin{align}
\parallel\vec{k}^\prime n\gg & = \frac{\sum_{\vec{G}^\prime} C^n_{\vec{k}^\prime+\vec{G}^\prime}|\vec{k}^\prime+\vec{G^\prime}\rangle}
                                    {\sqrt{\sum_{\vec{G}^\prime}|C^n_{\vec{k}^\prime+\vec{G}^\prime}|^2}}, 
\end{align}
which is obtained by renormalizing the plane-wave coefficients in Eq.~\ref{eq:regroup} for each $\vec{k}^\prime$. We can rewrite $|\vec{k}n\rangle$ as 
\begin{align}
|\vec{k}n\rangle & =  \sum_{\vec{k}^\prime} \parallel\vec{k}^\prime n\gg\ll\vec{k}^\prime n\parallel\vec{k}n\rangle. 
\end{align}
The inner product $\ll\vec{k}^\prime n\parallel\vec{k}n\rangle$ gives the probability amplitude of $|\vec{k}n\rangle$ to be found
in the conceptual state $\parallel\vec{k}^\prime n\gg$, and its square delivers the unfolded spectral weight for representing the band structure: 
\begin{align}
  |\ll\vec{k}^\prime n\parallel\vec{k}n\rangle|^2 = \sum_{\vec{G}^\prime}|C^n_{\vec{k}^\prime+\vec{G}^\prime}|^2, 
\label{eq:unfold}
\end{align}
where $\vec{k}$ can be referred to as the supercell crystal momentum for the case with the consideration of broken translational symmetry 
and $\vec{k}^\prime$ as the conceptual crystal momentum.

Several properties can be expected by representing the spectral weight of a supercell energy eigenstate in the introduced basis, $\parallel\vec{k}^\prime n\gg$,
which is what we mean by unfolding the band structure. For the case the supercell is a perfect supercell 
to a primitive unit cell, the unfolded weight can be either 0 or 1 for identifying the primitive-cell energy eigenstate. For the case the supercell is not perfect,
the unfolded weight is expected to be between 0 and 1 in reflecting the degree of symmetry breaking introduced to the primitive unit cell. For the case
the supercell is already a primitive unit cell that cannot be smaller, one can still choose a smaller conceptual unit cell to study the symmetry breaking
introduced to the chosen conceptual unit cell since there is no difference between the imperfect supercell and the primitive unit cell in the 
procedure for unfolding their bands into the Brillouin zones of the primitive and conceptual unit cells, respectively. Eq.~\ref{eq:regroup} also 
guarantees that folding the unfolded weight back to the supercell Brillouin zone would recover the full weight, 1, for each supercell eigenstate.  
More details of band-unfolding methods can be found elsewhere\cite{PhysRevB.71.115215,PhysRevLett.104.216401,
PhysRevLett.104.236403,PhysRevB.85.085201,PhysRevB.87.085322,PhysRevB.87.245430,lee2013unfolding,huang2014general,PhysRevB.90.115202,PhysRevB.90.195121}. 

\section{ARPES spectra}
\label{sec:arpes}

In this section, we will discuss the intensity measured by ARPES experiments and make connection to the unfolded spectral weight described in Sec.~\ref{sec:unfolding}.
The adopted approximations and the connection will be addressed in Sec.~\ref{sec:approximations} and Sec.~\ref{sec:connection}, respectively.

\subsection{ARPES intensity and approximations}
\label{sec:approximations}

The matrix element of describing ARPES spectra in the Coulomb gauge for
an initial state $|\psi_i\rangle$ to a final state $|\psi_f\rangle$ 
excited by a photon is $\langle\psi_f|\vec{A}\cdot\hat{\vec{P}}|\psi_i\rangle$ by the Fermi's golden rule
\cite{williams1977direct,RevModPhys.75.473,puschnig2015simulation,moser2016experimentalist}.
The $|\psi_i\rangle$ we consider is the energy eigenstate $|\vec{k}n\rangle$.
By inserting an identity operator composed of complete plane wave $|\vec{K}\rangle$, the measured momentum intensity $I$  
from $|\vec{k}n\rangle$ is proportional to:
\begin{align}
 & I(\vec{p},|\vec{k}n\rangle) \\
 \approx & |\langle\vec{p}|\psi_f\rangle|^2 |\langle\psi_f|\vec{A}\cdot\hat{\vec{P}}|\vec{k}n\rangle|^2 \nonumber \\
 = & |\langle\vec{p}|\psi_f\rangle|^2 |\sum_{\vec{K}} \langle\psi_f|\vec{A}\cdot\hat{\vec{P}}|\vec{K}\rangle\langle\vec{K}|\vec{k}n\rangle|^2 \nonumber \\
 \approx & |\langle\vec{p}|\psi_f\rangle|^2 |\sum_{\vec{K}} \vec{A}_{\vec{q}\rightarrow 0}\cdot \langle\psi_f|\hat{\vec{P}}|\vec{K}\rangle\langle\vec{K}|\vec{k}n\rangle|^2. 
\label{eq:arpes1}
\end{align}
In the optical limit, the photon wave vector $\vec{q}$ approaches zero and 
$\langle\psi_f|\hat{\vec{P}}|\vec{K}\rangle$ is assumed not to be altered by the vector potential $\vec{A}$. 
Since eventually what is measured is the momentum eigenstate $|\vec{p}\rangle$ in a constant vacuum potential, 
the measured intensity should be proportional to $|\langle\vec{p}|\psi_f\rangle|^2$. 
It is then common to assume 
\begin{align}
|\psi_f\rangle \approx |\vec{p}\rangle.
\end{align}
If the form of $|\vec{p}\rangle$ does represent the final state in dominating the physics so that the
integration can be performed over the all space with the same Born-von Karman boundary condition of the studied system,
the matrix element can be further simplified as
\begin{align}
 I(\vec{p},|\vec{k}n\rangle) \approx & |\sum_{\vec{K}} \delta_{\vec{p},\hbar\vec{K}} (\vec{A}_{\vec{q}\rightarrow 0}\cdot\vec{p}) \langle\vec{K}|\vec{k}n\rangle|^2 \nonumber\\
       = & |\vec{A}_{\vec{q}\rightarrow 0}\cdot\vec{p}|^2|\langle\vec{p}|\vec{k}n\rangle|^2,
\label{eq:arpes2}
\end{align}
which is equivalent to squaring the Fourier component $\langle\vec{K}=\vec{k}+\vec{G}|\vec{k}n\rangle$ because of $\delta_{\vec{p},\hbar\vec{K}}$.
This connects the photoelectron intensity with the momentum distribution of the initial state\cite{williams1977direct}.
Even though $|\psi_f\rangle$ is not replaced by $|\vec{p}\rangle$, the probability amplitudes delivered by the \textit{plane waves}\cite{PhysRevB.92.195148}, 
as shown in Eq.~\ref{eq:arpes1},
still contain abundant information because of the \textit{momentum detector}. 

\subsection{Connection between unfolded weight and ARPES intensity}
\label{sec:connection}
 
Let us make connection of the unfolded weight shown in Eq.~\ref{eq:unfold} to the photoelectron intensity presented in Eq.~\ref{eq:arpes2}, where the final state
is assumed to be $|\vec{p}\rangle$. 
By choosing the real-space conceptual cell as small as possible, the unfolded spectral weight of $|\vec{k}n\rangle$ can be considered as only being specified by $\vec{p}$ 
since the first Brillouin zone contains the entire space. Consequently, the weight exactly reproduces the factor $|\langle\vec{p}|\vec{k}n\rangle|^2$  
for describing the photoelectron intensity. However, a reasonable size of the conceptual cell should be chosen in order to simultaneously show the information 
of both photoelectron intensity and band structure of a \textit{periodic} system. Besides, the choice
of the conceptual cell could originate from theoretical interest rather than a fair comparison with ARPES spectra. More importantly, although the
overall strength of the Fourier components would decay with increasing kinetic energy of plane waves for a valence band, a converged periodic pattern could show up
in reflecting the effect of imposed periodic potential, which provides a natural choice of the conceptual cell. Finally, we mention that in the situation the connection 
cannot be easily made, other components in Eq.~\ref{eq:arpes1} or more complete description of ARPES spectra should be taken into account. 

\begin{figure*}[tbp]
\includegraphics[width=2.05\columnwidth,clip=true,angle=0]{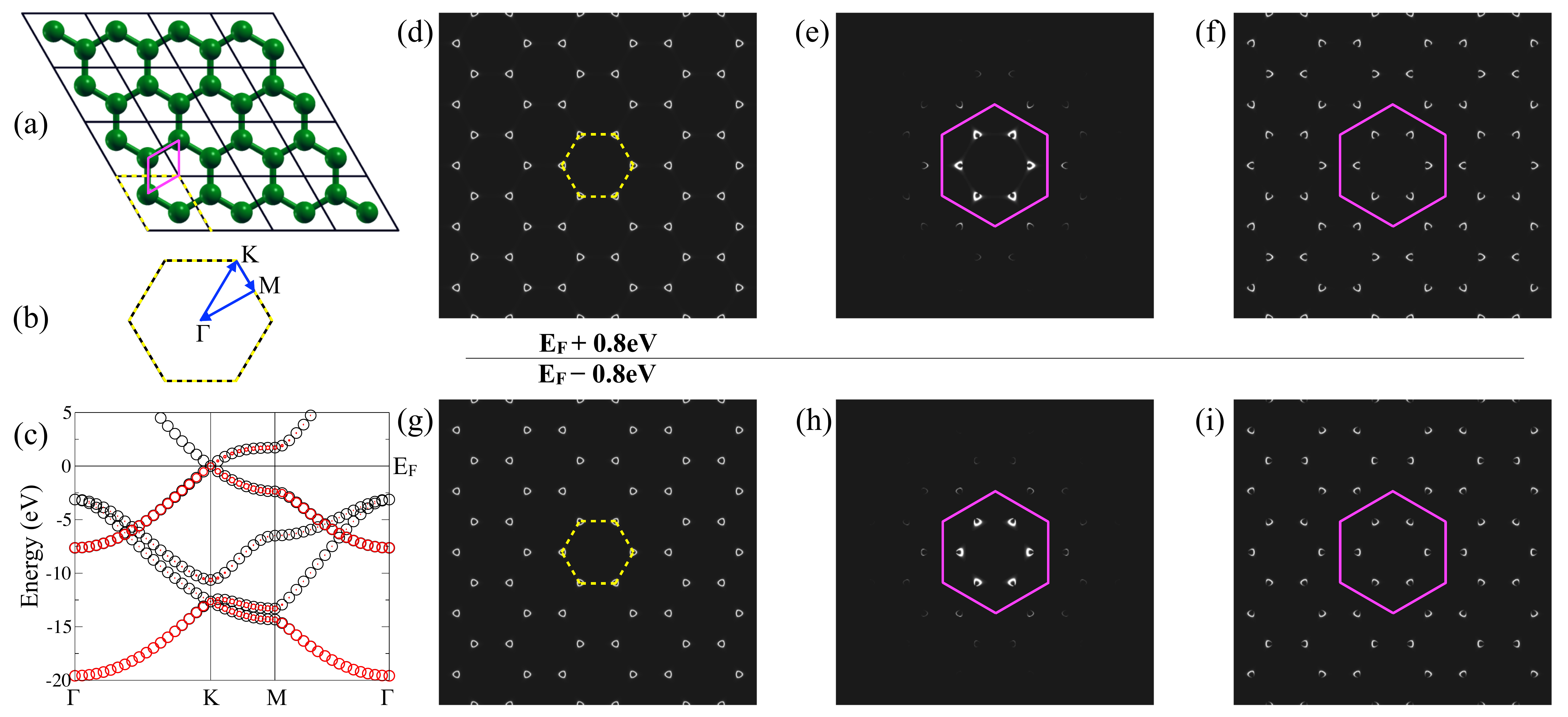}
\caption{
(a) Geometrical structure of graphene and (b) the corresponding first Brillouin zone. (c) The black and red circles show the spectral weight
represented in the Brillouin zones of primitive (dashed lines in (a)) and conceptual (magenta lines in (a)) unit cells, respectively, 
where the weight is represented by the diameter of a circle. (d) Constant-energy contour
at 0.8 eV above the Fermi level ($E_F$) in the primitive zone. (e) The same contour but represented by the Brillouin zone of $0.1\times0.1$ primitive unit cell. 
Regardless of the overall decreasing weight, the pattern follows the periodicity of the conceptual Brillouin zone (magenta lines). 
(f) The same contour represented in the conceptual Brillouin zone. (g),(h), and (i) are the same as (d),(e), and (f), respectively, but at 0.8 eV below $E_F$.
}
\label{fig:band}
\end{figure*}

\section{Dirac cones in graphene}
\label{sec:graphene}

We now discuss the electronic structure of graphene.
The computational details of first-principles calculations for graphene, including how to obtain the plane-wave coefficients from atomic-like orbitals, will
be discussed in Sec.~\ref{sec:details}. The analysis of the momentum-resolved band structure of Dirac cones in graphene using the unfolding method
will be given in Sec.~\ref{sec:cones}.

\subsection{Computational details}
\label{sec:details}

The first-principles calculations were performed using the OpenMX code\cite{openmx}. 
The norm-conserving pseudopotentials and optimized pseudo-atomic-orbital (PAO) basis were used\cite{Ozaki}.
The computationally efficient PAOs can benefit from the band unfolding methods using plane waves by
inserting plane waves into the following equation expressing $|\vec{k}n\rangle$ by $|\vec{k}M\rangle$, where $M$ denotes the orbital
and the other quantum numbers of the PAO:
\begin{align}
  |\vec{k}n\rangle & = \sum_{M} C^{PAO}_{\vec{k}n,M} |\vec{k}M\rangle \nonumber \\
                   & = \sum_{M} C^{PAO}_{\vec{k}n,M} \sum_{\vec{G}} |\vec{k}+\vec{G}\rangle\langle\vec{k}+\vec{G}|\vec{k}M\rangle \nonumber \\
                   & = \sum_{\vec{G}} (\sum_{M} C^{PAO}_{\vec{k}n,M} \langle\vec{k}+\vec{G}|\vec{k}M\rangle) |\vec{k}+\vec{G}\rangle.
\label{eq:lcao}
\end{align}
The advantages are that the completeness of plane waves is not essential 
for delivering acceptable spectral pattern in the post-processing calculations and the Fourier components are not restricted to the grids 
following the Born-von Karman boundary condition, allowing more freedom in choosing the conceptual cell. Here, 40 Ry was chosen for the plane-wave cutoff energy.
The generalized gradient approximation\cite{GGA} for the exchange-correlation functional was adopted. 
Two, two, and one optimized radial functions were allocated for the $s$, $p$, and $d$ orbitals for each carbon atom with a
cutoff radius of 7 Bohr. A cutoff energy of 300 Ry was used for numerical integrations and for the solution of the Poisson equation.
The $30\times30$ $k$-point sampling for the experimental lattice constant ($a$=2.4612\AA) was adopted in the calculation.

\subsection{Dirac cones in the one-carbon-atom zone}
\label{sec:cones}

The band structure of graphene is shown in Fig.~\ref{fig:band} (c) represented by the primitive and conceptual unit cells given in Fig.~\ref{fig:band} (a).
The choice of the conceptual cell containing one carbon atom becomes clear by looking at the momentum distribution of the massless Dirac fermions.
As an example of limiting case in the extended-zone representation, the unfolded spectral weight represented by the Brillouin zone of $0.1\times0.1$ primitive unit cell 
is shown in Figs.~\ref{fig:band} (e) and (h), which can be considered as the momentum distribution. 
Although the overall weight is decreasing, the periodic horseshoe pattern can be identified as following the periodicity of the one-carbon-atom conceptual cell in 
the reciprocal space, reflecting the effect of imposed periodic potential in the Hamiltonian to the massless Dirac fermions in graphene. 
For presenting both the spectra and the still informative \textit{band structure}, the natural choice of the conceptual Brillouin zone is the 
one-carbon-atom Brillouin zone that collects the weight of the same pattern into the first zone.  
With this choice, the band structure presented by the red circles in Fig.~\ref{fig:band} (c) and the constant-energy contours in Figs.~\ref{fig:band} (f) and (i)
give good agreement with ARPES experiments in comparison with the results shown in Figs.~\ref{fig:band} (d) and (g) as represented by the primitive zone.
The corresponding three-dimensional plot with the energy dependence can be found in Fig.~\ref{fig:cone}.
It should be noted that the periodicities revealed by Fourier components could differ 
in different bands. Generally speaking, different conceptual cells can be chosen for performing the unfolding depending on the purpose\cite{Kosugi}. 

\begin{figure}[tbp]
\includegraphics[width=0.90\columnwidth,clip=true,angle=0]{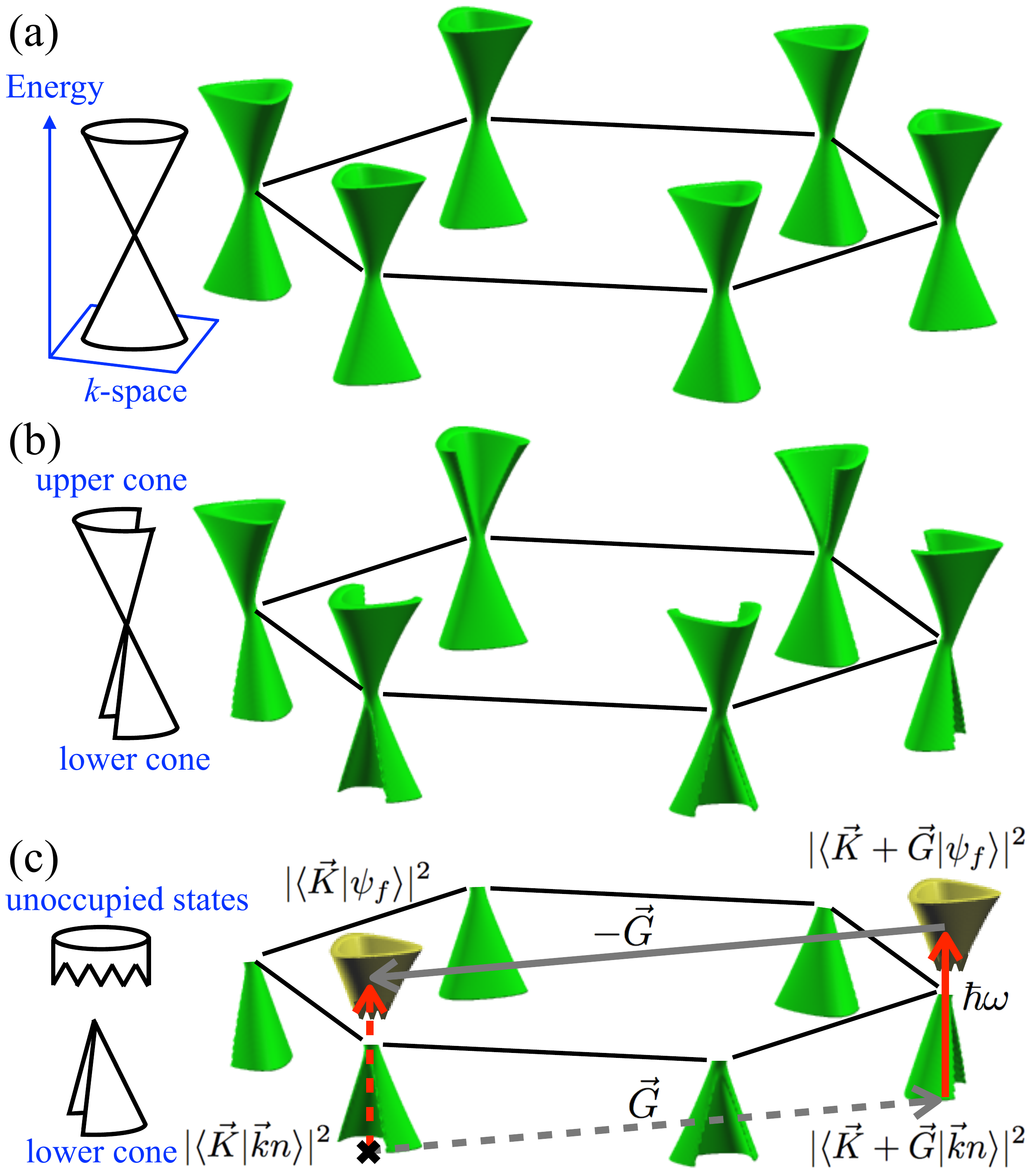}
\caption{
Dirac cones represented in the Brillouin zones of (a) primitive and (b) one-carbon-atom unit cells. 
(c) A virtual excitation process from occupied $|\vec{k}n\rangle$ to an assumed unoccupied final state $|\psi_f\rangle$ at
the decomposed $|\vec{K}\rangle\langle\vec{K}|\vec{k}n\rangle$ and $|\vec{K}\rangle\langle\vec{K}|\psi_f\rangle$ parts.
At first as assisted by the lattice with momentum $\hbar\vec{G}$, $|\vec{K}\rangle\langle\vec{K}|\vec{k}n\rangle$ reaches 
$|\vec{K}+\vec{G}\rangle\langle\vec{K}+\vec{G}|\vec{k}n\rangle$ 
and is then excited to $|\vec{K}+\vec{G}\rangle\langle\vec{K}+\vec{G}|\psi_f\rangle$ by an optical photon. 
Finally, it arrives $|\vec{K}\rangle\langle\vec{K}|\psi_f\rangle$ via $-\hbar\vec{G}$. The first step cannot actually function because of the negligible weight 
of $\langle\vec{K}|\vec{k}n\rangle$.
}
\label{fig:cone}
\end{figure}

Once the one-carbon-atom zone has been chosen, the unfolding procedure can also be performed without resorting to the plane waves, for example, by adopting 
the computationally more efficient methods using real-space Wannier functions or PAOs\cite{PhysRevLett.104.216401,lee2013unfolding}, to reveal spectral weight 
closer to the momentum distribution, such as the horseshoe patterns of the massless Dirac fermions in graphene\cite{lin2013create}.  
By adopting the one-carbon-atom unit cell, graphene is conceptually viewed as possession of periodic vacancies. The horseshoe patterns can then be 
understood from the interference between the same carbon atoms associated with different lattice vectors of the conceptual unit cell instead of considering 
two distinct carbon atoms in the primitive unit cell, suggesting that the Brillouin-zone-selection effect 
for the extinction of the intensity of Dirac cones\cite{PhysRevB.51.13614} in the momentum distribution
could also exist in a system having only one atom in the primitive unit cell. The form of the phase factor that 
can give the destructive interference can be found in the unfolding 
methods using real-space orbitals\cite{PhysRevLett.104.216401,lee2013unfolding}. Furthermore, the periodicity revealed by the momentum distribution,
such as the one-carbon-atom zone for the massless Dirac fermions in graphene, provides good 
guidance for experiments to seek for the investigated electrons other than the one provided by the reciprocal primitive vectors in general.

Note that $\langle\vec{p}|\vec{k}n\rangle$ and $\langle\vec{K}|\vec{k}n\rangle$ share the same mathematical form, 
$\int d\vec{r} e^{-i\vec{K}\cdot\vec{r}}\langle\vec{r}|\vec{k}n\rangle$, and the consistent Fourier components of 
massless Dirac fermions have been demonstrated in graphene studies\cite{PhysRevB.51.13614,bostwick2007quasiparticle,PhysRevB.77.195403,
PhysRevB.83.121408,PhysRevB.84.125422,PhysRevLett.107.166803,gierz2012graphene,hwang2014angle,hwang2015tight,PhysRevB.92.195148,puschnig2015simulation,moser2016experimentalist}, 
which were first analyzed by Shirley \textit{et al.}\cite{PhysRevB.51.13614} resulting from the interference between the $\pi$ orbitals of the two primitive-cell atoms that construct 
the energy eigenstates as given in Eq.~\ref{eq:lcao}, advocating the correctness of our calculations. 
Physically, instead of considering the plane wave as the final state, our study 
emphasizes the intrinsic properties of momentum distribution of the energy eigenstates, that is, the $\langle\vec{K}|\vec{k}n\rangle$ term 
in Eq.~\ref{eq:arpes1}, which is not related to the photon-excitation processes, 
such as how the photoexcited electrons reach the momentum detector from the two sub-lattices or the choice of the final state. 
Recall that the band structure in the extended-zone scheme is the information 
of the individual coefficient of $|\vec{k}n\rangle$ in each $|\vec{k}+\vec{G}\rangle$ basis. The unfolded weight, which is solely obtained from the initial state, 
sums over the squares of all relevant individual coefficients, $|\langle\vec{K}|\vec{k}n\rangle|^2$, and indicates the discontinuity of the Dirac cones in momentum space. 
This means that the momenta measured by ARPES experiments belonging to the missing pieces revealed by the unfolded weight cannot come from the massless Dirac fermions 
directly because of the closed excitation channels described in Eq.~\ref{eq:arpes1}. Following the discussions in Sec.~\ref{sec:arpes}, 
the momenta can only come from the unoccupied final states whose momentum distribution carries non-vanishing weight at those momenta.

To clarify the above point, we illustrate a virtual excitation process in Fig.~\ref{fig:cone} (c), where $|\vec{k}n\rangle$ carries no weight
at $|\vec{K}\rangle$ but some weight at $|\vec{K}+\vec{G}\rangle$. By impinging a photon with energy at $\hbar\omega$,
$|\vec{k}n\rangle$ could reach a symmetry-allowed final state via the $|\vec{K}+\vec{G}\rangle$ channel
but not via the $|\vec{K}\rangle$ channel because of the zero value at $|\langle\vec{K}|\vec{k}n\rangle|^2$. Given that
$|\langle\vec{K}+\vec{G}|\psi_f\rangle|^2$ and $|\langle\vec{K}|\psi_f\rangle|^2$ are both large,
the momentum intensity at $\vec{K}$ can then be measured from the contribution of $|\langle\vec{K}|\psi_f\rangle|^2$. 
Therefore, $|\vec{k}n\rangle$ carrying no weight at $|\vec{K}\rangle$ can still collapse to $|\vec{K}\rangle$. This does not 
violate momentum conservation since the lattice supplies the momentum $-\hbar\vec{G}$ in $|\psi_f\rangle$ via the periodic potential. It is possible that such a momentum 
cannot be supplied by the lattice via the periodic potential in a specific final state. 
The strong final-state dependence is consistent with the strong coupling with the photon energy found by the ARPES experiments\cite{gierz2012graphene,PhysRevB.92.195148}.

\section{Conclusion}
\label{sec:conclusion}

In conclusion, we provide alternative understanding of the massless Dirac fermions in graphene and also in other similar honeycomb materials via 
the broken Dirac cones revealed by the momentum-resolved band structure. 
The Dirac cones are shown to be intrinsically broken in momentum space beyond the periodic-zone scheme, and the revealed energy-momentum relationship allows 
easier analysis of the physical processes requiring the information of wave functions at each individual momentum, especially in coupling with physical quantities 
disobeying lattice translational symmetry, such as the momentum eigenstates measured by 
ARPES experiments. The revealed broken Dirac cones together with the momenta supplied by the lattice via the periodic potential,
where no phonon excitation is required,  
offer another point of view of the measured interplay of chirality and Berry phases between massless Dirac fermions and the final states that makes missing weight 
in Dirac cones retrievable. The retrieved momentum intensity is assisted with the properties of final states, not from the Dirac cones directly.
Although the physics we have discussed exists diversely in the literature, our study gives a comprehensive picture for the intrinsically
broken Dirac cones and clarifies several important issues in understanding the ARPES spectra.  
The broken Dirac cones together with the momenta supplied by the lattice with or without 
involving phonons and the tunability between complete and incomplete Dirac cones, including breaking rotational symmetry, via adatoms, substrates, or defects exhibit 
an interesting direction for designing advanced nanodevices. 

\setstretch{1.}
\section*{Acknowledgments}
This work was supported by Priority Issue (creation of new functional devices and high-performance materials to support next-generation industries) to be tackled by using Post `K' Computer, Ministry of Education, Culture, Sports, Science and Technology, Japan.

\bibliography{refs}

\end{document}